\pdfoutput=1
\documentclass[12pt]{iopart}

\usepackage{graphicx}
\usepackage{multirow}                         % when including figure files
\usepackage{multicol}
\usepackage{makeidx}         % allows index generation
\usepackage[bottom]{footmisc}% places footnotes at page bottom
\usepackage{hhline}
\usepackage{subfig}
\usepackage{color}
\usepackage{parskip}
\begin{document}

\title[Influence of Personal Preferences on Link Dynamics in Social Networks.]{Influence of Personal Preferences on Link Dynamics in Social Networks.}

\author{Ashwin Bahulkar$^{1}$, Boleslaw K. Szymanski$^{2}$}
\address{$^{1,2}$Rensselaer Polytechnic Institute, 110 8$^{th}$ Street, Troy, NY 12180}
\ead{$^{1}$bahula@rpi.edu,$^{2}$szymab@rpi.edu}

\author{Nitesh Chawla$^{1}$, Omar Lizardo$^{2}$}
\address{$^{1,2}$University of Notre Dame, Notre Dame, IN 46556}
\ead{$^{1}$nchawla@nd.edu,$^{2}$lizardo@nd.edu}

\author[cor1]{Kevin Chan,}
\address{US Army Research Laboratory, Adelphi, MD 20783}
\ead{kevin.s.chan.civ@mail.mil}

\begin{abstract}
We study a unique network data set including periodic surveys and electronic logs of dyadic contacts via smartphones. The participants were a sample of freshmen entering university in the Fall 2011. Their opinions on a variety of political and social issues and lists of activities on campus were regularly recorded at the beginning and end of each semester for the first three years of study. We identify a behavioral network defined by call and text data, and a cognitive network based on friendship nominations in ego-network surveys. Both networks are limited to study participants. Since a wide range of attributes on each node were collected in self-reports, we refer to these networks as attribute-rich networks. We study whether student preferences for certain attributes of friends can predict formation and dissolution of edges in both networks. We introduce a method for computing student preferences for different attributes which we use to predict link formation and dissolution. We then rank these attribute according to their importance for making predictions. We find that personal preferences, in particular political views and preferences for common activities help predict link formation and dissolution in both the behavioral and cognitive networks.
\end{abstract}

\maketitle

\section{Introduction}
\label{sec:1}
A key observation in the sociological literature is that persons have a preference to connect to others with similar attributes as themselves \cite{mcpher}. This propensity is usually referred to as ``homophily''.  Homophily can be based on a taste for similarity in values, beliefs, and attitudes. This is usually referred to as ``value homophily''. Homophily can also be based on a preference for similarity based on fixed or elective socio-demographic characteristics that define social groups (e.g. gender, age, race, social class); this is usually referred to as ``status homophily'' \cite{lazmert}. \textcolor{black}{Classic work in the social network analysis tradition sees homophily as a key tie formation mechanism \cite{rivera}. Other things being equal, we should expect that new connections between previously disconnected persons should more likely to emerge among those who share common attributes \cite{mcpher}}.

\textcolor{black}{While in some circumstances homophily operates as a direct tie formation mechanism, in some cases ties between similar alters may form not because people have a preference for people with similar attributes \cite{blau}, but because people with similar demographics or opinions end up participating in a common ``focus'' of activity such as groups or associations (such as a sports league or a cultural club) or common leisure activities (playing games, exercising) \cite{feld,fischer}. Thus, shared contexts or activities act as an additional tie formation mechanism, generating new connection among seemingly similar alters. Recent work examining the influence of shared contexts on tie formation processes confirms the impression that ties between similar alters are likely to be generated via this pathway \cite{mollenhorst1, mollenhorst2, mollenhorst3}}.

\textcolor{black}{Because either preferences for similar attributes or the sharing of common activity foci create non-random dependencies between node characteristics and the likelihood of tie formation and temporal persistence of social ties, both value and status homophily as well as shared contexts are key mechanisms implicated in explaining the temporal dynamics of social networks \cite{rivera}}. 

\textcolor{black}{Most previous work in social network analysis focuses on the mechanisms that generate new ties between previously disconnected nodes \cite{mcpher}. More recent work starting with \cite{burt} has focused on the phenomenon of \emph{tie decay}, which is the disappearance of an edge at a future point in time between two nodes that were previously connected \cite{rivera}. Although empirical work on tie decay continues to be relatively scarce, recent work has focused primarily on how \emph{structural} features of both the node (e.g. degree), and the edge (e.g. weight, triadic embeddedness) matter for decay \cite{raeder, hidalgo, martin}. This is primarily due to the fact that the usual data brought to bear to study decay processes in recent work (mostly based on networks constructed from interactions mediated via telecommunication technologies) is very thin on actual node attributes. Therefore very little is known as to how detailed node features such as attitudes, values, and leisure activities and shared contexts (but see \cite{lewis, vaisey}) influence tie decay processes. Given the fact that some of this works hints at the fact that shared attributes serve to delay the tie decay process \cite{burt}, it is likely that both status and value homophily as well as shared activities and foci may function as \emph{decay delaying} mechanisms, protecting ties from dissolution over time \cite{rivera,lewis}.}

To advance the study of the role of personal preferences on the dynamics governing the temporal evolution of social ties, we leverage a unique social science dataset, NetSense \cite{striegel} which is a dataset of about 200 students collected at the University of Notre Dame. The NetSense dataset is unique because, in contrast to other social network datasets, it is rich in attribute information: we know about every student's socio-demographic background, interests, opinions on social and political issues and the activities in which every student participates at multiple time points. We refer to this information as attributes of the students. Calls and messages exchanged between students are also recorded. In addition to this, students declare periodically who their top-twenty contacts are. \textcolor{black}{ NetSense data thus allows us to focus both on tie formation and tie decay processes defined over multiple (cognitive and behavioral) networks.} 

Using this information, we are able to identity two different social networks among the students, one is the behavioral network built from the call and message records in which students are the nodes and edges exist if a pair of students call or message each other. The other is the cognitive network built from the top twenty contacts reported by students in the periodic ego-network questionnaire. An edge between a pair of students exists if a student lists the other as a top contact in the surveys. Given the large amount of information that we have at our disposal about each student, we refer to both of these as attribute-rich networks. These networks are dynamic in nature, edges are created and dropped as persons add or subtract top contacts from the cognitive network and as communication volumes change over time in the behavioral network. 

In previous work using the same dataset, we examined how both value and status homophily as well as preferences for common activities affects the formation of ties. Consistent with sociological theories of homophily we found that, indeed, students with similar attributes are more likely to form ties with one another. In this previous work we used an aggregated count of the number of common preferences as the main predictor. One question that remains unanswered, therefore, is whether there is heterogeneity across attributes in terms of their importance in producing the homophily effect on tie formation. For instance, it is possible that political views of the other person may not matter much when it comes to forming friendship, but the shared activities in which the other person takes part might matter more. A different hypothesis is that the political views of the other person are of paramount importance, but only in the case of behavioral ties based on communication volume. 

Any ranking of which features matter more needs to be done while taking into account that different personal preferences matter more or less for different people. That is, we need to take into account that people may value different attributes more or less when making or breaking ties with other people, especially with regards to alternative values of the attribute. For instance, at each point in time, people have different values for an attribute, for example, the values a student can have for the attribute ``political orientation'' are conservative, moderate and liberal. People may have varying degrees of preferences for or against each of those values. We look at the following scenarios as the main motivation for our study: a person who is liberal may have a strong preference for liberals, but he may not have any preference for or against moderates and conservatives. While on the other hand, another liberal may have a strong preference against conservatives and a moderate preference against moderates. Our previous approach to understanding the role of homophily \cite{snaa}, capture neither of these scenarios. The simplifying assumption was that all liberals have the same preference for or against conservatives and moderates. The preference for or against an attribute value can be guessed from the neighborhood of a person in the network. A person having a strong preference for an attribute value would have higher than average number of friends with this attribute value, while a person with a preference against would have lower than average number of friends with this particular attribute value.

Our proposed method, which we call the \textit{Personal Preference Method}, takes these heterogeneous preferences into account and uses them to predict the formation and dissolution of edges. We look at the distribution of an attribute value in a person's neighborhood and compare it with the distribution of the attribute value in the entire network, and use this difference to measure the preference the person has for or against the particular attribute value. We further use these preferences combined with a machine learning approach to predict formation or dissolution of edges.

Our paper advances over previous work by extending the \textit{Personal Preference Method} to the task of predicting link dissolution. While prediction of edge formation has been studied well enough, prediction of edge dissolution has not received as much attention \cite{burt}. Predicting dissolution of edges is harder than predicting formation of edges since formation of edges is a more structured process than decay \cite{rivera}. We find significant differences in the number of attributes values over which nodes in edges which actually form agree, as compared to nodes in edges which do not ever form, with nodes forming edges agreeing across a wider range of attribute values a lot more than nodes which do not. However, differences between edges which dissolve and edges which persist are not very obvious \cite{snaa}. With the help of our preference based method, we are able to improve the performance of prediction for link dissolution significantly from the performance values mentioned in \cite{snaa}.

\section{Material and Methods}
\subsection{The NetSense data}
\label{sec:2}
% Always give a unique label
% and use \ref{<label>} for cross-references
% and \cite{<label>} for bibliographic references
% use \sectionmark{}
% to alter or adjust the section heading in the running head
The NetSense data used in this study consists of university students' reports listing their personal traits, interests, views and opinions on various social and political issues, and background at the beginning of every school semester from the Fall 2011 to the Spring of 2013 \cite{striegel}. At the beginning and end of each semester students are asked to fill out surveys where they list their friends. The data also consists of a record of the calls and texts exchanged between students participating in the study. We identify the evolving social networks among students out of this data.\\
\textbf{Call and text Messaging Data:}
We use the NetSense call and texts exchanged by students from August 2011 to May 2013. Each communication record consists of an entry for each call or text message, with the date, time, sender and receiver and duration or length of the communication. \\
\textbf{Friendship surveys:}
Each student can list up to 20 friends at the beginning of the semester. The friends are either survey participants, students on the campus not in the survey or family and friends outside the campus. We find that typically only two to three out of the 20 friends are survey participants. We form the network only out of the friendships which are between study participants.\\
\textbf{Node attributes:}
Students participating in the NetSense study filled out a survey at the beginning of each semester. Survey questions were about the studentsâ family background, major pursued in Notre-Dame University, activities on campus, their views on different social issues, and their political inclinations. All attributes have multiple possible values out of which a student selects one. For example, students can select if their political views are conservative or moderate or liberal. 
For each student we selected the following attributes from the NetSense data.\\\\
Student background
\begin{itemize}
\item Concentration of study/major
\item Family income
\item Race/ethnic identification (e.g. Black/White)
\item Religious affiliation (e.g. Catholic/Protestant)
\end{itemize}
Social and political views on:
\begin{itemize}
\item General political orientation (e.g. Liberal/Conservative)
\item Opinion about legality of abortion
\item Opinion about marijuana legalization
\item Opinion about homosexuality and the legalization of gay marriage 
\item Views on racial equality and affirmative action
\end{itemize}
Habits and Lifestyle:
\begin{itemize}
\item Drinking habits
\item Time spent weekly on activities like studying, partying, socializing, volunteering, campaigning for social causes, and exercising.
\end{itemize}
\textbf{Co-evolving networks in NetSense:}
From the NetSense data, we are able to identify and create two social networks among students. The first is the behavioral network, where an edge connects two students if calls are made or text messages are exchanged between them over the given semester. The second is the cognitive network, where an edge exists if one student lists the other as a top contact in the current survey. These two social networks evolve every semester, so we have four snapshots for both of the networks. The snapshots cover the following semesters: Fall 2011 semester ranging from August 2011 to December 2011, Spring 2012 semester lasting from January 2012 to May 2012, Fall 2012 semesters ranging from August 2012 to December 2012, and Spring 2013 semester lasting from January 2013 to May 2013. Since very few calls were made during the summer of 2012, we do not create a network for the summer semesters.  

\subsection{Related work}
\label{sec:rel}
Link prediction is a well-studied topic. The standard techniques for link prediction have been mentioned in \cite{lp} and \cite{lp1}. Most of the experiments there are on collaboration networks between researchers. However, none of the networks in these papers are as rich in node attributes as NetSense. We study how homophily in terms of node attributes affects link prediction in \cite{snaa}. While we were able to get reasonable results, the innovative \textit{Personal Preference Method} proposed in this paper improves the quality of link prediction in NetSense. 

Link dissolution is a less well-studied topic. Link dissolution in human mobility networks has been studied in \cite{mobil} and link persistence prediction has been studied on phone calling data in \cite{persis}. However, the networks studied there are not attribute-rich. \textcolor{black}{Several methods for analyzing the effect of different network properties like reciprocity of links, assortativity on formation and dissolution of links has been discussed in \cite{siena}. However, in contrast to the networks used in the study, the NetSense networks we study are much richer in node attributes. We also want to study the overall effect of all the node attributes, which is why we use the machine learning methods described below. We experimented with a maximum likelihood approach to predict links as mentioned in \cite{siena2} , however, we found that machine learning methods give much better performance. We also experimented with several statistical methods and found that machine learning methods give us the best predictions. } We studied link dissolution in the NetSense networks in \cite{snaa}, however, the results were not very encouraging. With our new \textit{Personal Preference Method}, we are able to make significant improvements over the previous effort. In addition, in \cite{snaa} the experiments were performed exclusively on the behavioral network, in this paper we extend the enhanced approaches to dynamics of decay in the cognitive network.

\subsection{A case for the Personal Preference Method}
\begin{figure}[!tbp] 

  \centering
  \subfloat[Semester 1]{\includegraphics[width=0.5\textwidth]{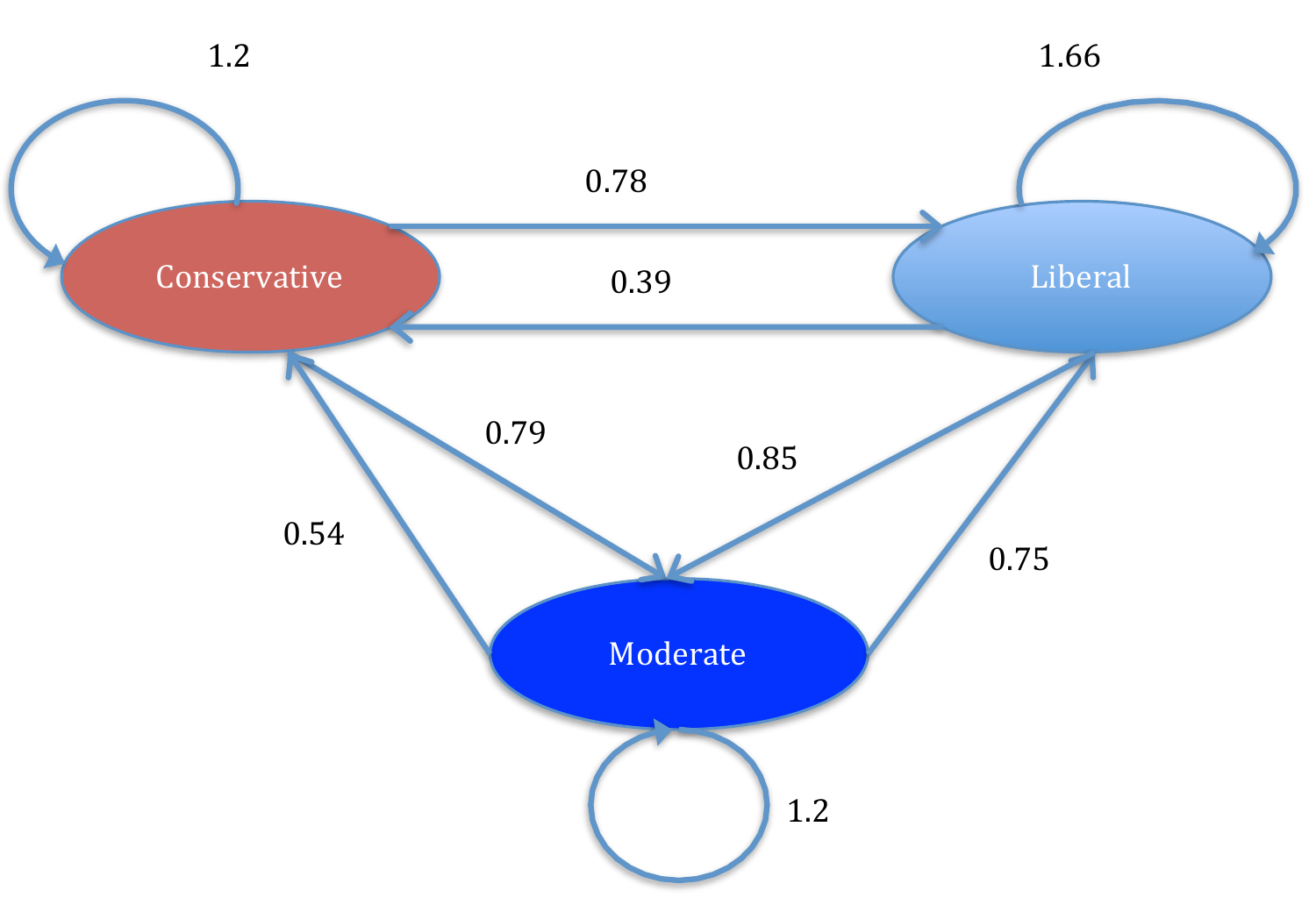}\label{fig:1}}
  \hfill
  \subfloat[Semester 2]{\includegraphics[width=0.5\textwidth]{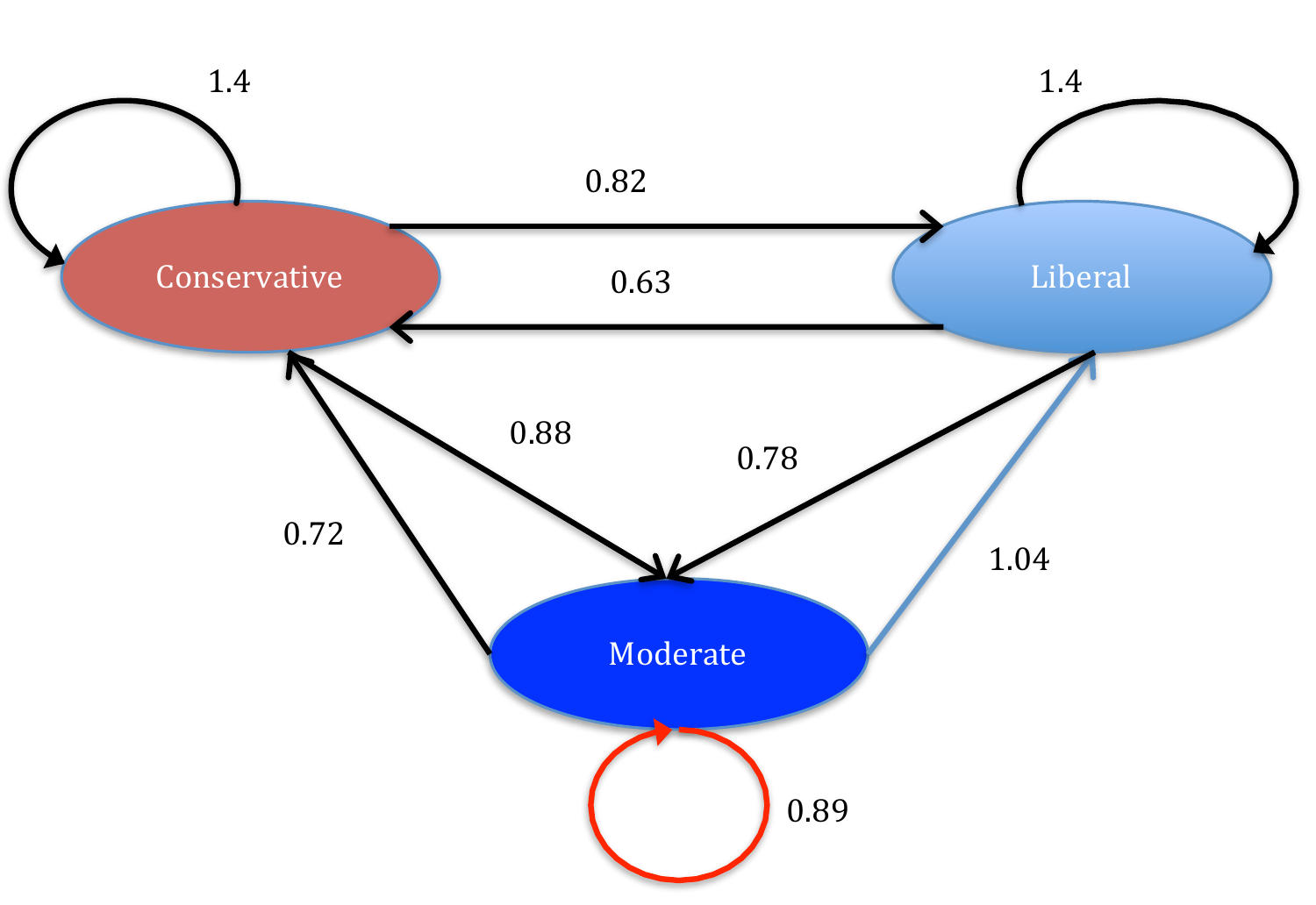}\label{fig:2}}\\
  \subfloat[Semester 3]{\includegraphics[width=0.5\textwidth]{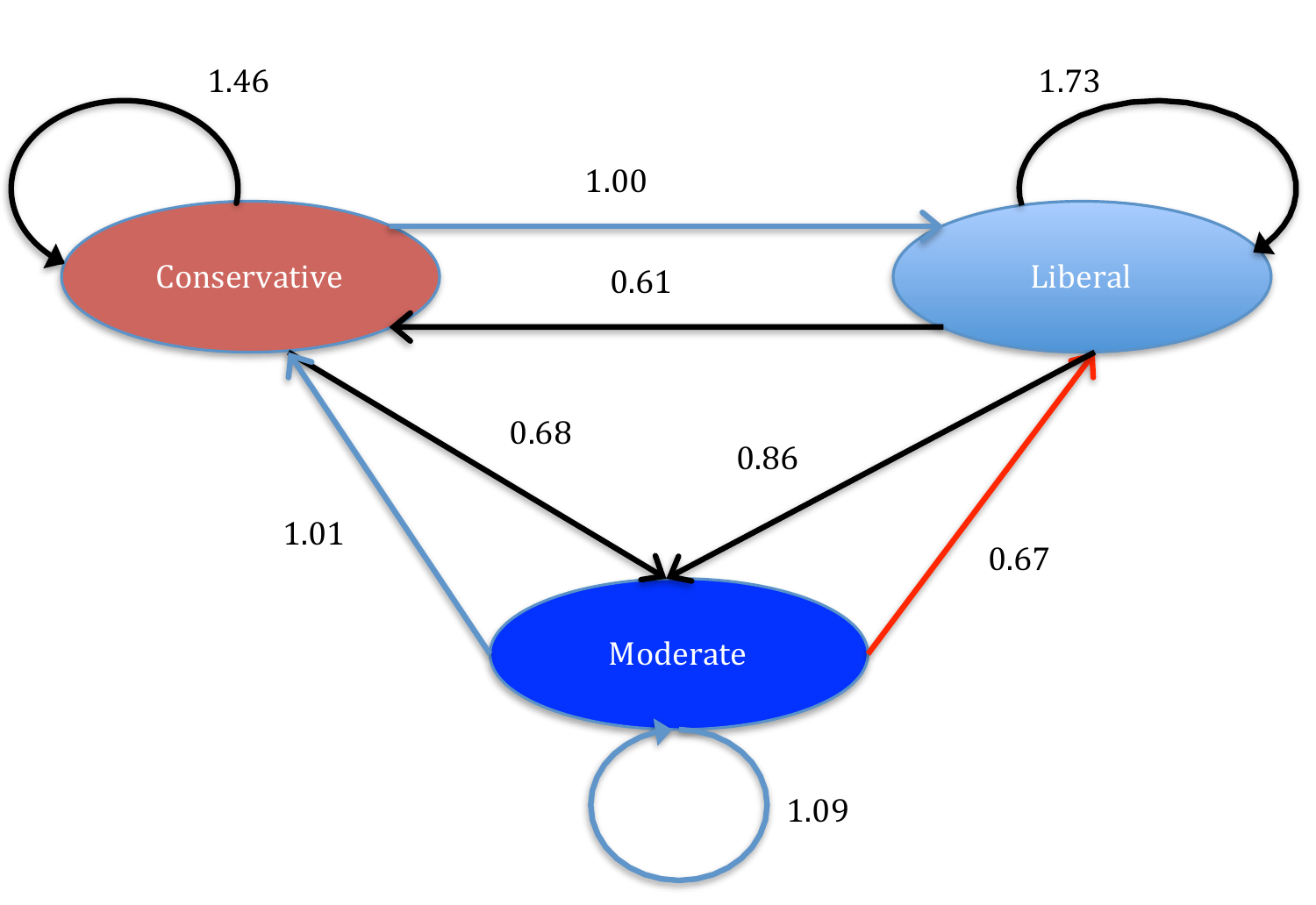}\label{fig:3}}
  \hfill
  \subfloat[Semester 4]{\includegraphics[width=0.5\textwidth]{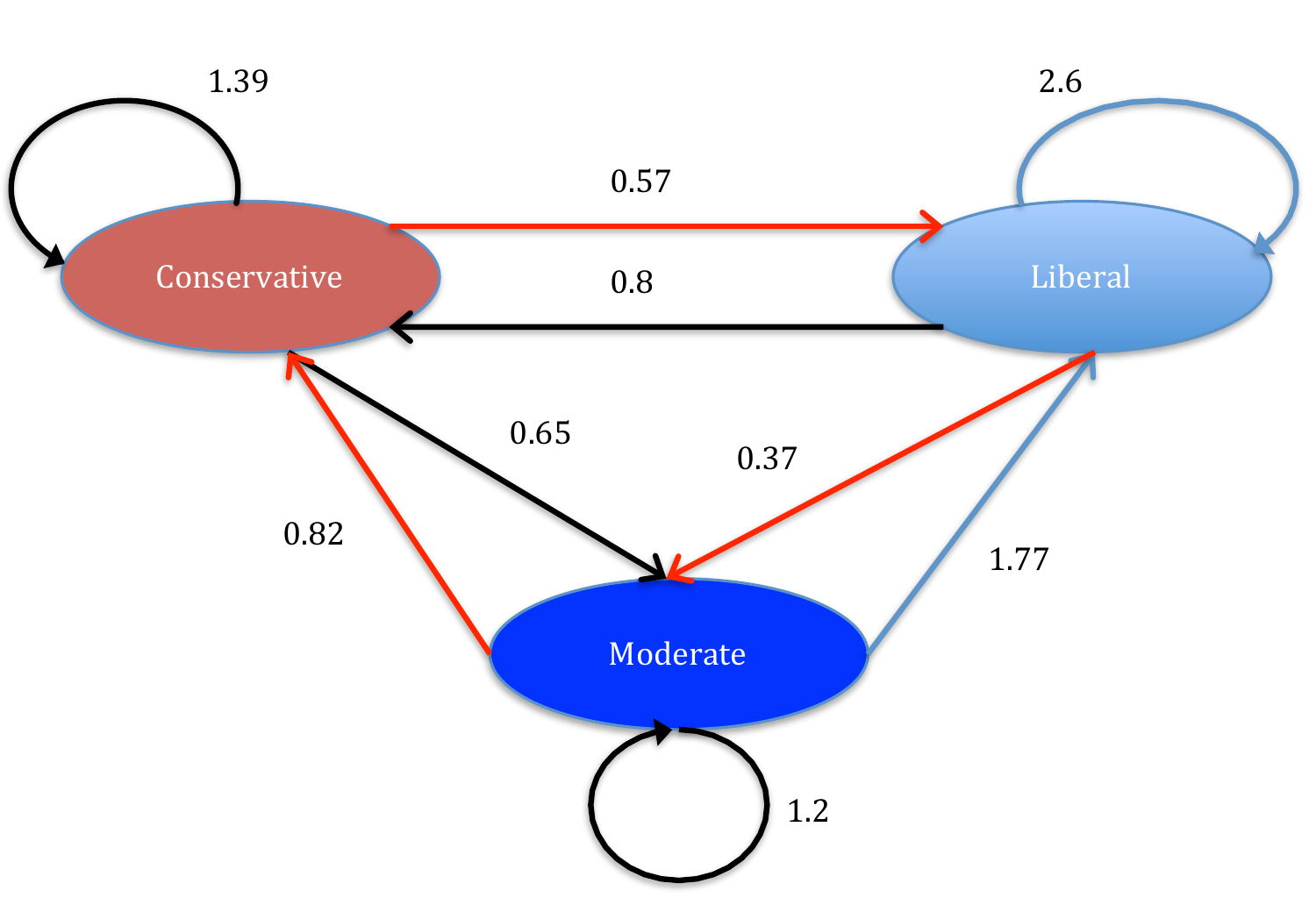}\label{fig:4}}
  \caption{Views on politics, the average preference of nodes with a particular attribute value for all values of the attribute.}
  \label{figpol}
\end{figure}

\begin{figure}[!tbp] 

  \centering
  \subfloat[Semester 1]{\includegraphics[width=0.5\textwidth]{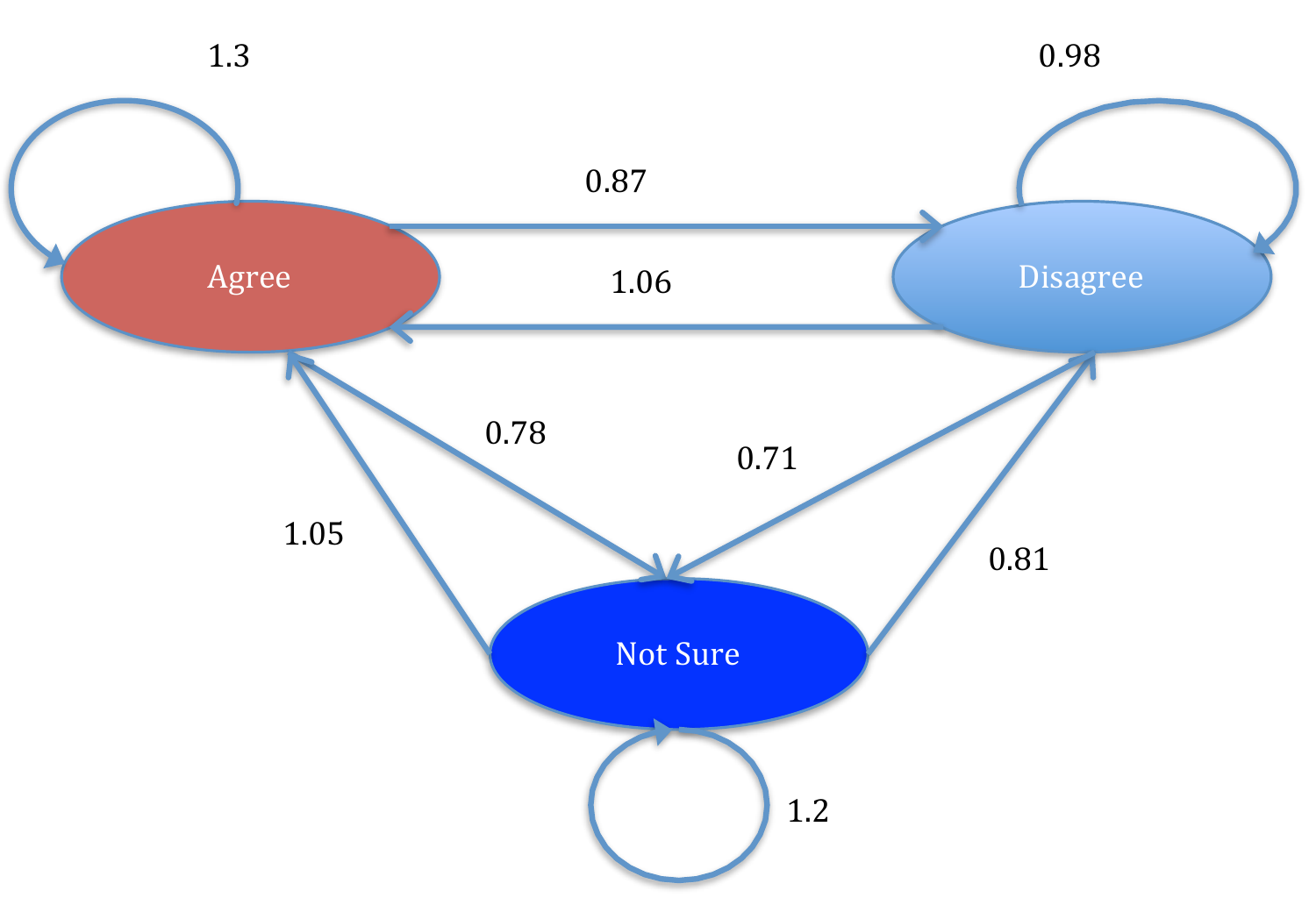}\label{fig:1a}}
  \hfill
  \subfloat[Semester 2]{\includegraphics[width=0.5\textwidth]{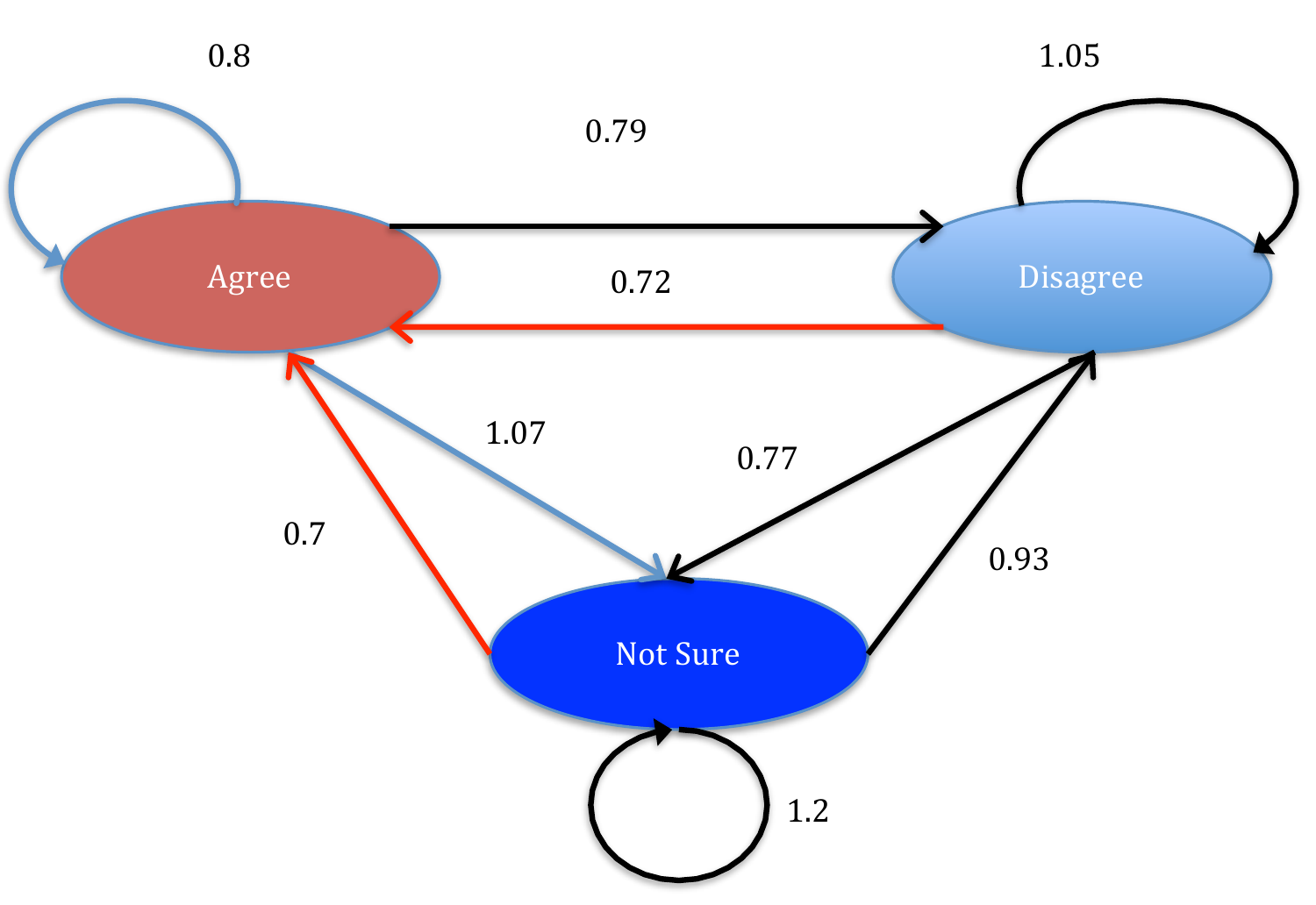}\label{fig:2a}}\\
  \subfloat[Semester 3]{\includegraphics[width=0.5\textwidth]{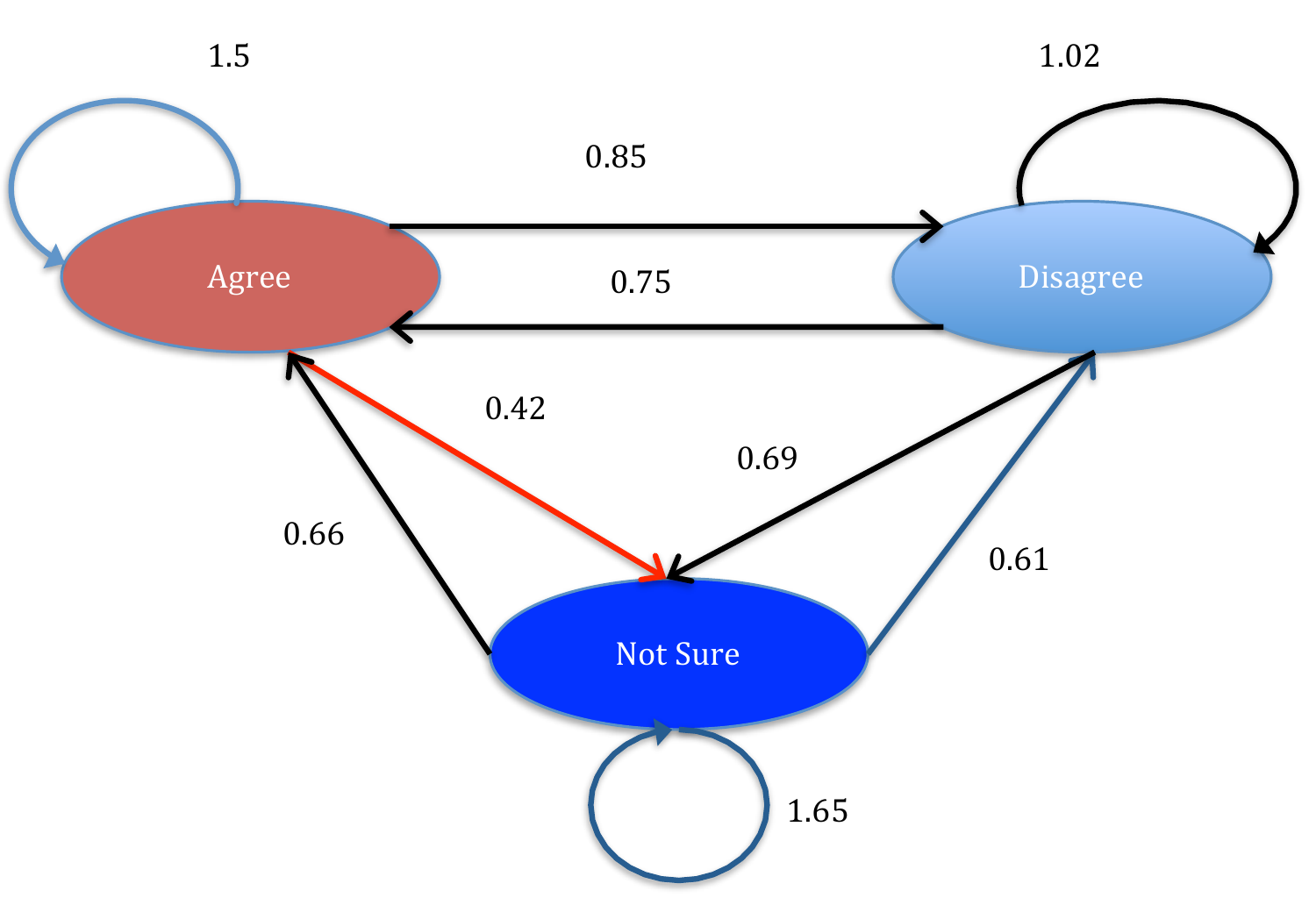}\label{fig:3a}}
  \hfill
  \subfloat[Semester 4]{\includegraphics[width=0.5\textwidth]{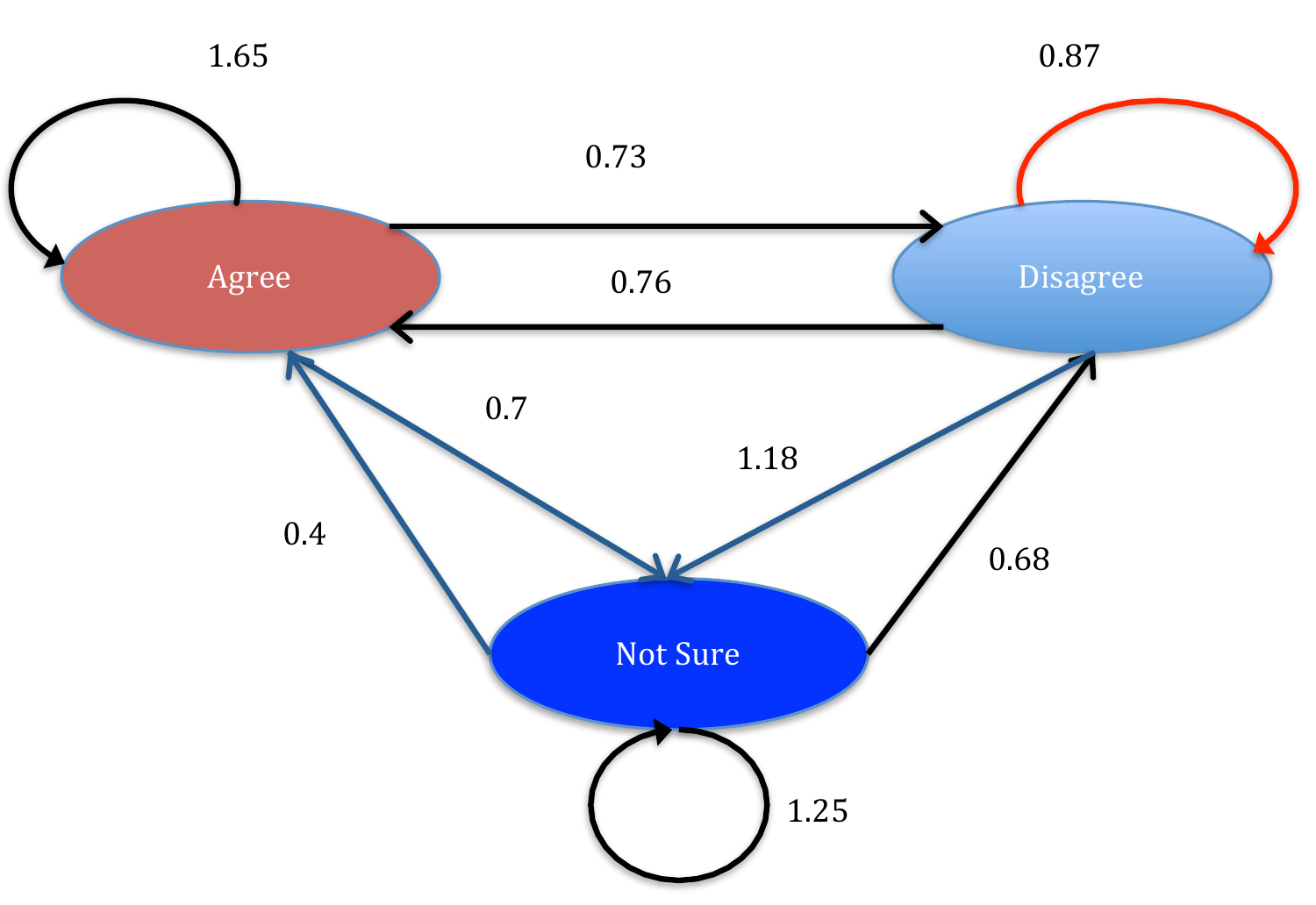}\label{fig:4a}}
  \caption{Views on gay marriage legalization, the average preference of nodes with a particular value for all values of the attribute.}
  \label{figgy}
\end{figure}

\begin{figure}[!tbp]
  \centering
  
  \includegraphics[width=0.5\textwidth]{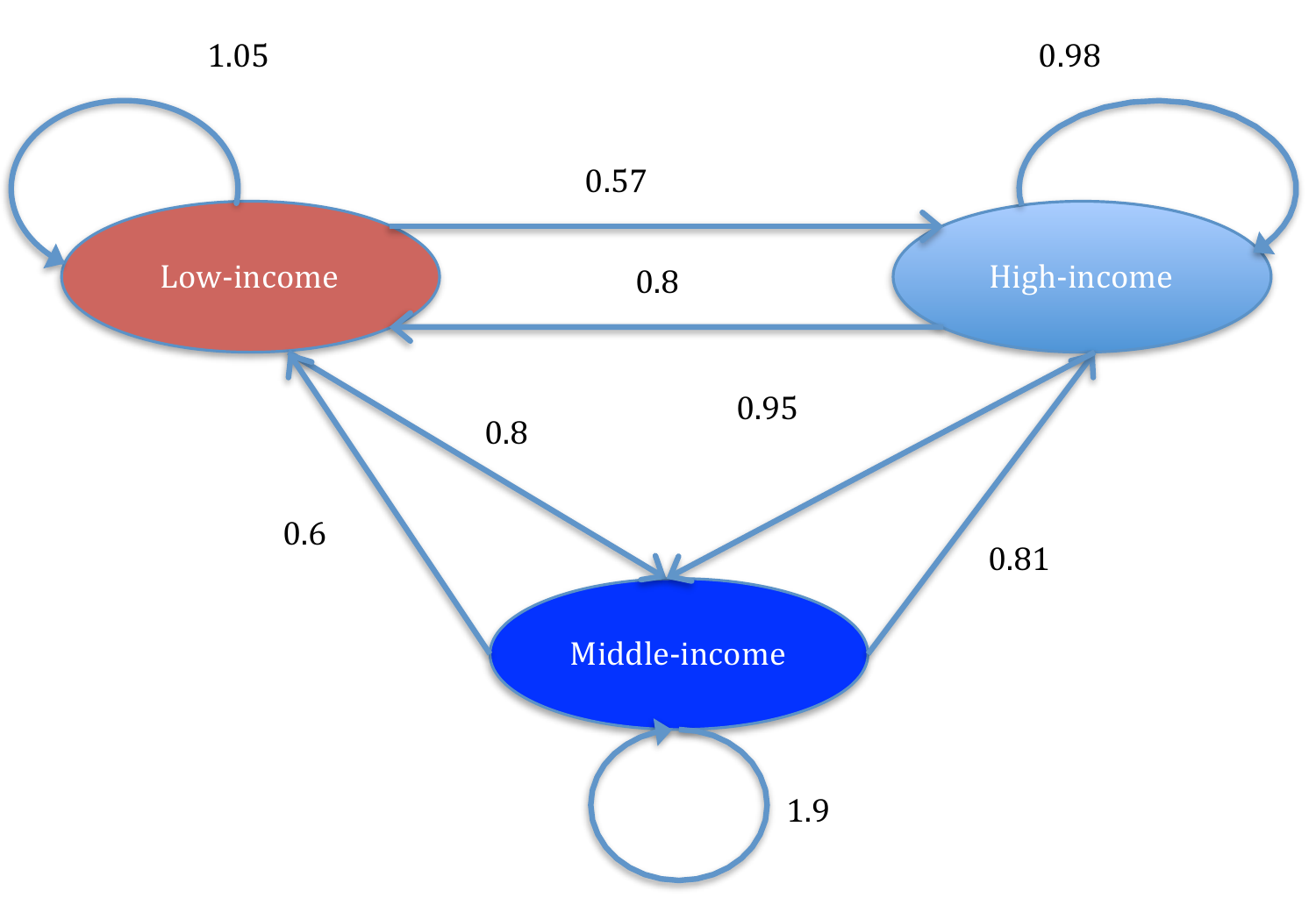}
  \setlength\belowcaptionskip{-8pt}
  \caption{Parental income, the average preference of nodes with a particular value for all values of the attribute. Unlike others, parental income data was collected only once by the first survey.}
   \label{fig:income}
\end{figure}

In our work in \cite{snaa}, we made the (reasonable) assumption, grounded in the sociological literature, that people prefer to form friendships with people who are exactly like them and we performed link prediction based on this assumption. In this section, we look at the limitations of this assumption. In the previous paper, our link prediction algorithm had assumed that a liberal would prefer liberals the most, followed first by moderates, and then by conservatives. Now, the distance between the preference values of moderates and liberals with each other and between moderates and conservatives was assumed to be equal. Also, we assumed that all liberals would have the same preferences. Preliminary analyses on the NetSense data reveal that this was probably too simplistic of an assumption. \textcolor{black}{To illustrate dynamics of the links between people with different attribute values, in Figure \ref{figpol}, Figure \ref{figgy} and Figure \ref{fig:income}, we visualize changes in the strength of connection between them. Attributes values are linked by the preferences between nodes possessing them. A value greater than 1 means a higher than average preference, a value less than 1 signifies a lower than average preference, a value around 1 means an average preference. A blue line stands for a significant increase in the preference from the previous semester, a red line means a decrease in the preference and a black line means no significant change in the preference. From these three figures, we observe the average preferences held by all the nodes possessing a particular attribute value, for all the values of this attribute. Although we find consistently that while people often have a strong preference for other people with the same attribute value as theirs, the preference for people with other attribute values does not necessarily follow a predetermined order. For example, in Figure \ref{figpol} one would expect that liberals would have a higher preference for moderates than conservatives, but this is not always what we observe. Also, the changing preference values over the semesters makes us reconsider our previous assumption that all the persons possessing a particular value for an attribute have the same preferences. This leads us to propose a method where we can account for every person's preferences and use them to make predictions as to which attributes are more important for social network evolution.}

\subsection{Methodology}
We want to find out how well node preferences for different attributes predict the formation and dissolution of edges in both the cognitive and behavioral networks. We also want to know which attributes play the most important role in the formation and dissolution of edges. We first introduce a technique where we measure the preference of every node for every attribute value, based on the neighborhood of the node. We then propose a method to convert these preferences of a pair of nodes into features, which are used to predict the formation or dissolution of edges using machine learning. We then consider the different machine learning techniques used to make these predictions. Finally, we elaborate on how we obtain the relative importance of every attribute in the process of making predictions. 

\subsubsection{Link formation prediction}
Link formation prediction can be seen as a classification problem where we predict whether an edge which connects a pair of nodes, but still has not been formed, will form in the future or not \cite{lp-desc}. We use machine learning techniques to predict the formation of edges. Features for this prediction task are the preferences of the two nodes to be connected by the edge for each other's values of each of the attributes along with a network topology-based feature. We describe the \textit{Personal Preference Method}, which creates these features, in Section \ref{sec:rel}. The network topology based feature we use is the number of common neighbors between edge endpoints, or, in other words, the number of neighbors shared by those two nodes.\\
\textbf{Classification task and the training and test datasets.}
New edges are predicted by machine learning models that learn from the way new edges were created in the past. To perform link formation prediction, we need three successive snapshots of the network, say \textit{net1}, \textit{net2} and \textit{net3}. We predict which nodes will be joined by an edge in \textit{net3}. So, yet to be formed edges in \textit{net2} are the test set. Out of these edges, the ones which are formed in \textit{net3} are true positive, and ones which are not formed in \textit{net3} are true negative examples. The machine learning model has to be trained on the past, so we use the first two networks to train the model. The unformed edges in \textit{net1} are the training set. Out of these, the ones which are created in \textit{net2} are positive examples while the ones which do not form in the \textit{net2} are negative examples.\\
\textbf{Unbalanced classification}
The link formation prediction classification task is an unbalanced one. With \textit{n} nodes and \textit{e} edges in the network, there are $\left(\frac{n^2-1}{2}\right)- e$, possible edges which usually is many times more than there are new edges formed. We observe that most nodes tend to link to other nodes when they are separated by no more than three hops. So, in our dataset, we consider only the edges whose nodes are separated by at the most three hops. However, still there are 50 timer more negative examples than 
the positive ones. With machine learning algorithms, we have observed that we often need to compromise either on accuracy or recall. Recall measures the percentage of new edges that were predicted to be such, while accuracy measures the fraction of the predictions, regardless if they were positive or negative, that were accurate. The goal of our classification task is to select a prediction model that gives us the best balance between accuracy and recall. We choose a model which identifies as many new edges as possible, while at the same time not classifying too many negative edges as positive. We use the ROC curve with a weight assigned to recall being five times larger than that assigned to accuracy and choose the best model accordingly.

\subsubsection{Link dissolution prediction}
Link dissolution prediction too can be seen as a classification problem, where we predict whether an edge will dissolve or not. In \cite{snaa}, we found that predicting dissolution did not yield very good results. We know that a decrease in communication is a strong signal of declining friendship as shown in \cite{previous16}. Yet, the cognitive edges are binary, they either exist or not and the change comes once a semester. In contrast, the behavioral edges have weights, so considering these facts, there are cognitive edges that may be under the process of dissolution, but this may not necessarily be reflected in the current friendship survey. To capture the process of dissolution, we define dissolving communication edges as those which reduce their communication in the succeeding semester to a third of their existing communication volume. Moreover, the cognitive edges whose corresponding communication edge is dissolving are also classified as dissolving. For clarity, an edge that get dissolved in the current  semester is also considered dissolving (this is by definition, for communication edges, but not necessarily for cognitive edges). With this definition, we redefine the classification task to predict dissolving edges and not only edges to be dissolved. To measure the weight of a communication edge, we aggregate the number of calls and text messages, assigning the weight 10 to each call and weight 1 to each message corresponding to the typical ratio of the number of messages to the number of calls in our data. We use machine learning techniques to predict the dissolving edges. The features used for prediction here are the same as those used for link formation prediction.\\\\
\textbf{Test and training sets}
We perform link dissolution prediction using machine learning models that learn from edges the have been dissolving in the past. The classification task is very similar to that of link formation prediction. For the classification task, we have three successive networks, \textit{net1}, \textit{net2} and \textit{net3}. We want to predict which the edges existing in \textit{net2} will be dissolving in \textit{net3}. From the nodes and network structure in \textit{net2}, we predict which edges would be dissolving in \textit{net3}. The machine learning model learns from the edges existing in \textit{net1} and dissolving in \textit{net2}. So, edges in \textit{net1} form the training set. Edges which are dissolving in \textit{net2} are true positive, while edges which are not dissolving in the \textit{net2} are true negative examples. Similarly, for edges in \textit{net2} that form the test set, edges which are dissolving in \textit{net3} are positive examples, and edges which are not are negative examples. 

\subsubsection{Machine learning techniques used}
We use the standard Support Vector Machines (SVM), Linear Regression, and k-Nearest Neighbors (k-NN), Random Forests and Naive Bayes classifiers as classification algorithms for all the classification tasks. They are the most commonly used classifiers in several link prediction works such as \cite{lp1}, \cite{lp} and \cite{lp-desc}.\\ 

\subsubsection{Validation set}
The validation set is used to fine-tune parameters of the machine learning 
algorithms. These parameters differ from algorithm to algorithm. In linear regression, we need to select the best threshold, with SVM and also with linear regression, we need to decide whether to use higher order features, with k-NN we need to select the best value of $'k'$ and with random forests, we need to select the best number of trees. The validation set contains randomly selected 20\% of the training set. 

\subsubsection{Computing node preferences} \label{pref-sec}
We compute preference of a node for every possible value of every attribute a node can have. The preference a node has for a particular attribute value is computed based on how different the percentage of the node's friends having the particular attribute value is from the percentage of all the nodes having that particular attribute value in the entire network. This percentage of friends being lower than that percentage for the entire network indicates a negative node bias towards  that particular attribute value, while opposite relation of these percentages indicates a preference for this attribute value. These percentages being equal indicates node's neutral attitude towards that particular attribute value. We use the statistical Z-Score~\cite{zscore} to measure how far from average is the number of friends with the said attribute for the given node. Z-Score is expressed in the standard deviation units. We normalized the Z-Score values into the range of [0, 1] using the Z-Score tables~\cite{zscore}.\\
\\
\textbf{The Personal Preference Method to compute node preferences}\\
\textbf{Input}: A set \textit{Attrset} of attributes, with each attribute $a$ in the set having a set of possible values $a.values$ that a node could have.
Each node \textit{n} in the network has a set of neighbors \textit{Nset}. Every node has a value for all the attributes.\\
\textbf{Output}: For every node \textit{n} in the network, for each value \textit{v} of each attribute, a preference value \textit{n.Preference(v)} between 0 and 1 is returned, with 1 denoting a strong positive preference, 0 denoting a strong negative preference, and a value of 0.5 indicating no preference. \\
\\
\textbf{Step 1: Calculating the distribution of each attribute value in the network.}\\ 
For each attribute \textit{a} from \textit{Attrset}\\
\hspace*{6mm} For each value of attribute $a$ \textit{v} $\in$ \textit{a.values}, find the percentage of nodes with the value \textit{v}. We refer to this value as $a.v.percentage$. \\
\\
\textbf{Step 2: Calculating the preferences for each node}\\
For each node $n$ in the network\\
\hspace*{4mm}For each attribute \textit{a} from \textit{Attrset}\\
\hspace*{8mm}For each value \textit{v} of the attribute \textit{a} from $a.values$\\
    	\hspace*{8mm}1. Calculate the Z-Score:\\
        \hspace*{12mm}Z-Score=($x$-$\mu$)/$\sigma$, as defined in \cite{zscore}.\\
        \hspace*{12mm}where, $x$ is the actual number of friends of $n$ with the particular attribute value.\\
        \hspace*{12mm}$\mu$ is the expected number of friends with the particular attribute value in the network, which is $n.noNeighbors \times a.v.percentage$, where $a.v.percentage$ is obtained in step 1 and \textit{n.noNeighbors} is the total number of neighbors $n$ has.\\
        \hspace*{12mm}$\sigma$ is the standard deviation.\\
        \hspace*{8mm}2. Convert the Z-Score to a normalized range between 0 and 1 using the Z-Score table, and assign it to $n.Preference(v)$. \\
        
\subsubsection{Building the features for classification}
For the machine learning task, our feature set is computed from the preference scores of the nodes of the edge for every attribute. We define two methods here, one where preferences of both nodes matter, so we multiply the preference of one node by that of another for that particular attribute. We call this method the \textit{Equal Preference Method}. Another method considers only the lower of the preference values of the two nodes. We want to find out if this simpler approach results in good predictions. We call this method the \textit{Minimum Preference Method}.\\
\textbf{Method for converting node preferences to edge attributes}\\
\textbf{Input:} Edge $e$ with nodes $n1$ and $n2$. For a node $n$, the preference for a value $val$ of each attribute a is denoted by  $n.Preference(a.val)$. \\
\textbf{Output:} For each of the attributes, an agreement value between $n1$ and $n2$ is calculated. For the edge $e$, agreement on an attribute $a$ is denoted by $e.Agreement(a)$.\\
\textbf{Method:}\\
	For each attribute $a$\\
    \hspace*{4mm}$n1$'s value for $a$ is denoted by $a.val1$, $n2$'s value for $a$ is referred to as $a.val2$. \\
    \hspace*{4mm}\textit{Equal Preference Method}:\\
    \hspace*{4mm}$e.Agreement(a) = n1.Preference(a.val2) \cdot n2.Preference(a.val1)$\\
    \hspace*{4mm}\textit{Minimum Preference Method}:\\
    \hspace*{4mm}$e.Agreement(a) = Min(n1.Preference(a.val2),n2.Preference(a.val1))$\\
We use these feature values for classification of edges.

\subsubsection{Estimation of attribute importance}
We estimate the relative importance of every attribute that leads to formation and dissolution of edges in both the networks. We look at the coefficients of every attribute in the function returned by the linear regression classifier and use them to estimate the relative importance of each attribute. A higher coefficient is associated with higher importance in classification. The values of the coefficients cannot be interpreted literally, since several dependencies exist among the features. However, the coefficients are still a fair indicator of how important each attribute is, and the ranking of these weights still gives us a fair idea of the relative importance of each attribute.
\section{Results and Discussion}
\subsection{Link formation prediction results}
We measure link prediction performance using accuracy and recall. Recall measures the fraction of the created edges that were predicted as such while accuracy measures the fraction of predictions that were correct. We find that the accuracy and recall rates for link formation prediction have significantly improved over our earlier approach presented in \cite{snaa}. This clearly demonstrates the benefit of using the node preferences for attribute values. 

\begin{figure}[!tbp]
  \centering
  \subfloat[ROC curve for behavioral network]{\includegraphics[width=0.5\textwidth]{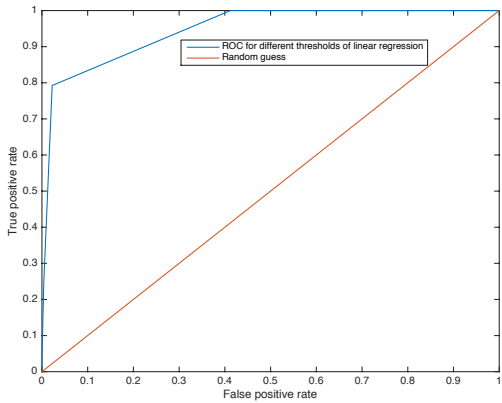}\label{fig:roc1}}
  \hfill
  \subfloat[ROC curve for cognitive network]{\includegraphics[width=0.5\textwidth]{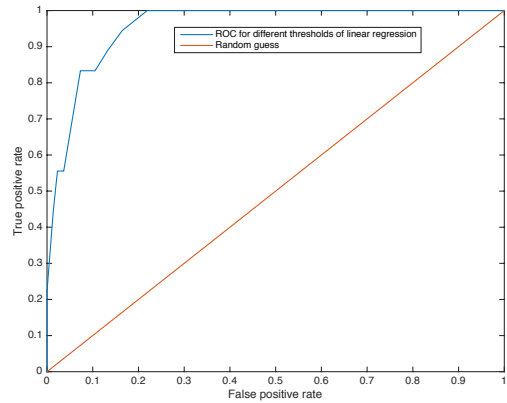}\label{fig:roc2}}
  \caption{\textcolor{black}{The plot of the ROC curve for prediction of edge formation in both the behavioral and cognitive networks for different thresholds of edge weight in the second semester (the curves for other semesters are similar). The curve shows that the models fit the data well, but that higher the recall, higher the false positive rate. The prediction in behavioral network performs a little better than it does in the cognitive network}}
\end{figure}

Linear Regression, SVM and k-NN classifiers yield the best results, with high accuracy and high recall. Random Forests and Naive Bayes classifiers performed poorly. In \cite{snaa}, the best recall and accuracy achieved were 76\% each while here both were above 97\%. Table \ref{tab:lpb} lists the results for the behavioral network, which is the same network we had used in \cite{snaa}. We omitted results for Naive Bayes classifier due to its poor performance, We report results only for the \textit{Equal Preference Method}, since \textit{Minimum Preference Method} performed slightly lower. Table \ref{tab:lpc} shows the results for the cognitive network, which we did not analyzed in \cite{snaa}. We make predictions for formation of links in the third and the fourth semester. The recall rates for the cognitive network are significantly higher than those for the behavioral network, possibly because of a smaller network size and a stronger tendency among nodes to adhere to their preferences while forming cognitive friends, as opposed to forming edges in the behavioral network. \textcolor{black}{We also present the ROC curves in Figure \ref{fig:roc1} and Figure \ref{fig:roc2}.} 

In \cite{snaa}, we had used Singular Value Decomposition(SVD) for feature extraction. While this had enabled us to get an increase in the recall value then, using SVD with our \textit{Personal Preference Method} here did not make any difference. We had also used additional features for classification like the "number of attributes on which the nodes of an edge agree" but using this feature did not make a difference in the results of our classification.
\begin{table}[h]
\centering
\caption{Link formation prediction results for the behavioral network}
\label{tab:lpb}       % Give a unique label
%
% Follow this input for your own table layout
%
\begin{tabular}{c|c|c|c}
\hline
\multicolumn{2}{c}{Semester} \vline &Semester 3  &Semester 4 \\
\hline
%Semester 1  &71   &76  &61  &84\\

\multirow{2}{*}{Linear Regression} &Accuracy &97.5 &97.8\\
&Recall &79.5 &92.8\\
\hline
\multirow{2}{*}{SVM} &Accuracy &96.6 &96.5\\
&Recall &89.9 &92.8\\
\hline
\multirow{2}{*}{k-NN} &Accuracy &96.5 &97.8\\
&Recall &92.8 &88.1\\
\hline
\multirow{2}{*}{Random Forests} &Accuracy &98.5 &98.0\\
&Recall &38.8 &58.1\\
\hline
\end{tabular}
\end{table}

\begin{table}[h]
\centering
\caption{Link prediction results for the cognitive network}
\label{tab:lpc}       % Give a unique label
%
% Follow this input for your own table layout
%
\begin{tabular}{c|c|c|c}
\hline
\multicolumn{2}{c}{Semester} \vline &Semester 3  &Semester 4 \\
\hline
%Semester 1  &71   &76  &61  &84\\

\multirow{2}{*}{Linear Regression} &Accuracy &74.6 &90\\
&Recall &100 &100\\
\hline
\multirow{2}{*}{SVM} &Accuracy &92.8 &77.8\\
&Recall &83.6 &93.3\\
\hline
\multirow{2}{*}{k-NN} &Accuracy &94.5 &90.0\\
&Recall &88.9 &94.5\\
\hline
\multirow{2}{*}{Random Forests} &Accuracy &94.5 &89.9\\
&Recall &58.1 &50.4\\
\hline
\end{tabular}
\end{table}

\iffalse
\begin{table}[h]
\centering
\caption{Link Prediction results for the cognitive network}
\label{tab:lpc}       % Give a unique label
%
% Follow this input for your own table layout
%
\begin{tabular}{c|c|c|c|c|c|c|c|c}
\hline
\multirow{2}{*}{Semester} & \multicolumn{2}{c}{Regression}  & \multicolumn{2}{c}{SVM} & \multicolumn{2}{c}{k-NN} & \multicolumn{2}{c}{Random Forest}\\
\hhline{~--------}
     & Accuracy &Recall &Accuracy &Recall & Accuracy &Recall &Accuracy &Recall\\
    
\hline
%Semester 1  &71   &76  &61  &84\\
Semester 3  &74.6	&100	&92.8	&83.6 &94.5	&88.9 &94.5	&58.1\\
Semester 4  &90	&100	&77.8	&93.3 &90	&94.5 &89.9	&50.4\\

\hline
\end{tabular}
\end{table}
\fi

\subsection{Link dissolution prediction}
To assess the performance of link dissolution prediction, we measure the precision along with accuracy and recall, since this classification task is balanced, unlike link formation prediction. Precision is defined as the fraction of edges predicted to be dissolving by the classifier that are actually dissolving. Interestingly, the performance of the prediction of to be dissolved edges was very similar. We found that the accuracy, recall and precision rates for prediction of edge to be dissolved have significantly improved over our earlier method presented in \cite{snaa}. We are able to predict a significantly larger fraction of to be dissolved edges than in the past. This demonstrates the benefit of using the node preferences.

We also benefit from using edges which do not dissolve but whose nodes reduce the communication volume between them significantly. Linear Regression gives us the best results, with high accuracy, recall and precision. SVM too yields good results, but not as good as Linear Regression. Random forests and k-NN tend to yield pretty low recall results. We do not report the results for Naive Bayes, k-NN and Random Forests classifier, since their accuracy rates were much lower than for the remaining methods. %The performance values for edges which completely dissolve is slightly higher than these values. 
Table \ref{tab:ldc} lists results for both the behavioral and cognitive networks. We combine the results of two semesters for the cognitive network since there are very few edges which dissolve in the fourth semester. We report results from the \textit{Equal Preference Method}. 
%\textcolor{black}{We also show the ROC curves in Figures \ref{fig:roc1-diss} and \ref{fig:roc2-diss}.} For comparison, the best recall and accuracy achieved in \cite{snaa} with dissolution were 69\% and 63\%, respectively. \\

\begin{table}[h]
\centering
\caption{Link Dissolution Prediction results for the behavioral and cognitive networks using Linear Regression}

\label{tab:ldc}       % Give a unique label
%
% Follow this input for your own table layout
%
\begin{tabular}{c|c|c|c}
\hline
Semester &Precision &Recall &Accuracy\\

\hline
%Semester 1  &71   &76  &61  &84\\
Semester 3  &80.6	&92.5	&83.1	\\
Semester 4  &82.4	&90.1	&84.2	\\
\hline							
Cognitive Network	\\
\hline
Semester 3 and 4  &81.2	&80.1	&75.2	\\
\hline
\end{tabular}
\end{table}

%\begin{figure}[!tbp]
%  \centering
%  \subfloat[ROC curve for behavioral network]{\includegraphics[width=0.5\textwidth]{roc-diss-behav}\label{fig:roc1-diss}}
%  \hfill
%  \subfloat[ROC curve for cognitive network]{\includegraphics[width=0.5\textwidth]{roc-diss-cognitive}\label{fig:roc2-diss}}
%  \caption{\textcolor{black}{The plot of the ROC curve for prediction of edge dissolution in both the behavioral and cognitive networks for different thresholds of edge weight for the second semester (the curves for other semesters are similar). The curve shows that the models fit the data well, but that higher the recall, higher the false positive rate. The prediction performance is higher for the behavioral communication network than for the cognitive network.}}
%\end{figure}

\subsection{Relative importance of attributes}
We look at the coefficients of all attributes returned by linear regression. Table \ref{tab:weight} lists the normalized relative weights of all the attributes, while Table \ref{tab:weight-rank} lists the rank of every attribute in the classification. A higher relative weight implies higher importance during classification. We present these rankings for the predictions for the fourth semester for both edge formation and dissolution prediction for both the networks. We observe that the coefficient weights are highly correlated for both semesters for the behavioral network, so the results shown here are just from the predictions made for semester 4.

We find that political views ranks high when it comes to formation of friendships in both networks and it ranks high in dissolution of edges in the behavioral network as well. Parental income, number of common neighbors and time spent on common activities such as volunteering, and exercising seem to rank higher in the formation of edges than in the dissolution of edges. Common activities such as partying and camping appear to matter more for dissolution than they do for link formation. Drinking habits, views on abortion, college major, race, and religion seem to rank low in all the networks, for both formation and dissolution. Views on the legality of gay marriage ranks higher in formation than in dissolution especially in the cognitive network for which it is the first and third most important feature, respectively. Views on moral propriety of homosexuality, time spent in clubs and socializing seem to rank higher in dissolution than in formation of edges.

\begin{figure}[!tbp]
	
  \centering
  
  \includegraphics[width=0.75\textwidth]{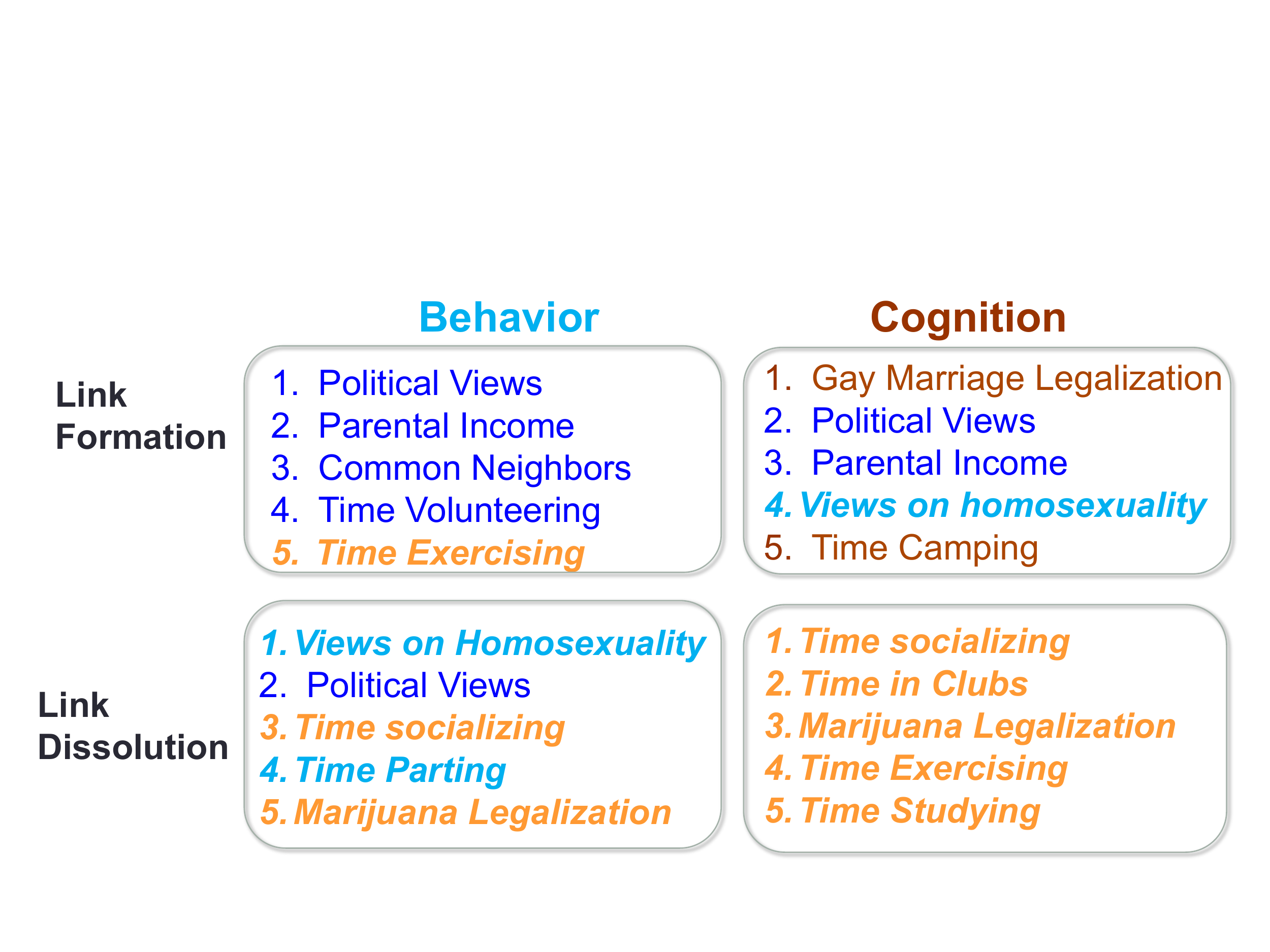}
  \setlength\belowcaptionskip{-14pt} 
  
  \caption{Comparison of top ranking attributes for link formation and dissolution of behavioral and cognitive edges. The four categories of edge dynamics, going from left to right and from top to bottom are: behavioral link formation, cognitive link formation, behavioral link dissolution and cognitive link dissolution. The five highest ranking attributes for each category are shown in different colors. Interestingly, dissolution of cognitive edges has its attributes highest ranked among all categories. Only political views are shared by three categories, while parental income is common for link formation, time socializing and marijuana legalization are common for link dissolution, while time exercising is shared diagonally and views on homosexuality are common along anti-diagonal. Of the 13 distinct attributes listed, seven are unique for one category: common neighbors, time volunteering, gay marriage legalization, time camping, time parting, time in clubs and time studying. More than half of listed attributes, seven, are some form of time spent together but interestingly, different forms of spending time together have impact on different link categories of formation or dissolution.}
\label{fig:attrfig}
\end{figure}

Comparing Figure \ref{fig:attrfig} with Table \ref{tab:weight-rank}, we can observe that six attributes: parental income, gay marriage legalization, political views, time volunteering, time in clubs, and time studying, are among top 11 attributes for all four categories of edge dynamics. These attributes have highest influence on link formation and dissolution of both cognitive and behavioral edges. Among them, political views attribute is the most potent, being the first for behavioral link formation, the second for cognitive link formation and behavioral link dissolution and 11 for cognitive edge dissolution. Collectively some form of spending time together is also very important; time socializing is ranked high for link dissolution, it is the first for cognitive and the third for behavioral link dissolution, while time volunteering is ranked fourth for behavioral link prediction with time camping and time parting ranked fifth and sixth for formation of cognitive links.

\begin{table}[h]
\centering
\caption{Weights of different attributes}
\label{tab:weight}       % Give a unique label
%
% Follow this input for your own table layout
%
%\begin{tabular}{l|c|c|c|c}
\begin{tabular}{p{6.2cm}| p{1.8cm}|p{1.7cm}|p{1.8cm}|p{1.7cm}}
\hline
\multirow{2}{*}{Semester} & \multicolumn{2}{c}{Link Prediction} \vline & \multicolumn{2}{c}{Link Dissolution}\\
\hhline{~----}
     &Behavioral &Cognitive &Behavioral &Cognitive\\

\hline
%Semester 1  &71   &76  &61  &84\\
Political views  &1	&0.6	&0.78 &0.22\\
Parental income &0.95	&0.58	&0.26	&0.27\\
Drinking habits &0.22	&0.05	&0.13 &0.08	\\
Views on abortion &0.35	&0.02	&0.12 &0.15	\\
Views on gay marriage legalization &0.75	&1	&0.25 &0.67	\\
Views on homosexuality &0.2	&0.55	&1	&0.28\\
Views on marijuana legalization &0.16	&0.25	&0.44 &0.12\\
Major &0.38 &0.18	&0.21	&0.26\\
Race &0.42	&0.17	&0.12	&0.21\\
Religion &0.33	&0.15	&0.09	&0.11\\
Number Common neighbors &0.84 &0.35 &0.13 &0.07\\
Time spent on volunteering &0.8 &0.35 &0.14 &0.48\\
Time spent on exercising &0.76 &0.15 &0.02 &0.54\\
Time spent on studying &0.69 &0.33 &0.21 &0.49\\
Time spent on partying &0.35 &0.42 &0.55 &0.11\\
Time spent on university clubs &0.49 &0.23 &0.26 &0.92\\
Time spent on socializing &0.71 &0.12 &0.57 &1\\
Time spent on camping &0.69 &0.43 &0.01 &0.35\\
\hline
\end{tabular}
\end{table}

\begin{table}[h]
\centering
\caption{Ranks of different attributes}
\label{tab:weight-rank}       % Give a unique label
%
% Follow this input for your own table layout
%
\begin{tabular}{p{6.2cm}| p{1.8cm}|p{1.7cm}|p{1.8cm}|p{1.7cm}}
\hline
\multirow{2}{*}{Semester} & \multicolumn{2}{c}{Link Prediction} \vline & \multicolumn{2}{c}{Link Dissolution}\\
\hhline{~----}
     &Behavioral &Cognitive &Behavioral &Cognitive\\

\hline
Political views  &1	&2	&2 &11\\
Parental income &2	&3	&6	&8\\
Drinking habits &16	&17	&13 &17	\\
Views on abortion &14	&18	&15 &13	\\
Views on gay marriage legalization &6	&1	&8 &3	\\
Views on homosexuality &17	&4	&1	&9\\
Views on marijuana legalization &18	&10	&5 &14\\
Major &12 &12	&10	&10\\
Race &11	&13	&14	&12\\
Religion &15	&14	&15	&16\\
Number Common neighbors &3 &7 &12 &18\\
Time spent on volunteering &4 &8 &11 &6\\
Time spent on exercising &5 &16 &17 &4\\
Time spent on studying &9 &9 &9 &5\\
Time spent on partying &13 &6 &4 &15\\
Time spent on university clubs &10 &11 &7 &2\\
Time spent on socializing &7 &15 &3 &1\\
Time spent on camping &8 &5 &18 &7\\
\hline
\end{tabular}
\end{table}

\section{Long term changes in the network}
\textcolor{black}{
We define strong edges as those whose nodes agree on more than $t_s$ fraction of their attributes, and the remaining edges are called weak. Then, we check if generally the strong edges have a higher chance of survival than the weak edges do and how the difference depends on the threshold $t_s$ used to define strong edges. Experimentally, we found that $t_s=0.75$ is the best value for separating strong edges from the week ones. With this threshold, 80\% of strong edges survive in semester 1, compared to 44\% of weak ones. In semester 2, the survival rates are 63\% for strong and 55\% for weak edges, while in the third semester, these rates are 75\% for strong versus 78\% for weak edges, a slight reverse in the trend. At the same time, the average fraction of strong versus weak edges changes slightly from 17\% to 14\% to 21\% and finally to 20\% over the four semesters. These semester-to-semester changes are not consistent because there are weak edges being created and dissolved in the network all the time.
However, if only the edges which are at least one semester old are considered, then the clear trend is uncovered showing a steady increase of fraction of strong edges. This fraction grows from 17\% in the second semester to 21\% in the third and to 26\% in the fourth semester. So over time, some tendency emerges to stabilize and increase homophily of the surviving edges.}

\section{Conclusions}
Using the user's preferences for different attribute values we are able to make high quality predictions of formation and dissolution of edges. We have shown that this method is able to increase the performance of predictions in the NetSense networks in comparison to other strategies used in the past. We believe this method would be useful for making predictions in other attribute-rich networks and demonstrates how preferences of nodes can be harnessed to predict formation and dissolution of edges and thus contribute to our understanding of behavioral dynamics in social systems \cite{rivera}. We also identified the relative importance of all the attributes in the formation and dissolution of edges. We found different attributes being top ranked for formation and dissolution of edges, suggesting that different factors might be responsible for formation and dissolution of ties between people. We also found attributes occasionally having different rankings for predictions in behavioral and cognitive networks, suggesting that different factors play a role in the formation and dissolution of social ties based on subjective importance versus those based on behavioral frequency.

In all, the results are consistent with and provide important extensions of sociological approaches that see value homophily as a form of ``cultural matching'' and an important mechanism in both the formation of new social ties \cite{lizardo,vaisey} and dissolution of existing ones. \textcolor{black}{Value homophily mechanism (e.g. views on homosexuality and gay marriage) seems particularly important when it comes to predicting the formation of new links in the cognitive network (based on subjective prominence) and when predicting behavioral dissolution of links.} 

This is particularly salient in the fact that political views emerged as the only factor that modulates both link formation and dissolution in the two networks (see Figure \ref{fig:attrfig}). This is in line with recent work \cite{macy} on the increasing salience of politics and the link between political views and lifestyles as an amplification mechanism (via homophily and social influence) driving patterns of social and geographical segregation in contemporary societies \cite{diprete}. \textcolor{black}{The fact that self-placement in the liberal-conservative continuum emerged as a preponderant predictor even when considering other attitudes and values associated with political orientation (e.g. views on abortion, marijuana legalization, homosexuality) seems to indicate that persons are sorting into homogeneous group based on their self-identification as ``conservative'' or ``liberal.'' This seems consistent with political views serving as marker of social identity \cite{socialID}, and not just as a factor impacting values and attitudes.} 

In addition, the preponderant role of common activities in generating link dissolution in the cognitive network (see Fig. \ref{fig:attrfig}) lends support to Feld's theory of social foci \cite{feld} as an important complement to the value homophily mechanism in patterning tie decay in social networks \cite{burt}. Essentially, this means that once students stop having a set of common activities bringing them together for interaction, they decline in terms of subjective prominence as a ``top contact.'' Note that in this respect, the mechanisms that produce new cognitive links are of a different nature than those that account for their decay (see upper right box of Figure \ref{fig:attrfig}) \cite{rivera}. 

Finally, the relative lack of importance of group-level identifications (``status homophily'') in modulating the temporal evolution of social ties in this network (e.g. common identities based on gender, race, and religion) is consistent with the view that most of the matching observed along these lines is modulated via either cultural matching or common-activity mechanisms \cite{vaisey,wimmer}. In all, the results reported here provide important sociological advances in our understanding of the role of cultural processes in generating patterns of connectivity and segregation in human social networks. 

\section{Declarations}

%\end{Ethics approval and consent to participate}
\textbf{Conflict of interests}
The authors declare that there is no conflict of interest regarding the publication of this article\\
\textbf{Acknowledgments and funding}
We would like to thank Panagiotis Karampourniotis for discussions. This work was supported in part by the Army Research Laboratory under Cooperative Agreement Number
W911NF-09-2-0053 (the Network Science CTA), and by the Office of Naval Research (ONR) grant
no. N00014-15-1-2640. The views and conclusions contained in this document are those of the
authors.




%\bibitem{roc} Bradley, Andrew P.: The use of the area under the ROC curve in the evaluation of machine learning algorithms. Pattern Recogn., 1145--1159 (1997)

\end{document}